\documentclass[aps,prb,twocolumn,superscriptaddress, show pacs]{revtex4-1}
\usepackage{bm, amsmath}
\usepackage{graphicx}
\usepackage{color}
\usepackage{epstopdf}
\usepackage{graphicx}
\usepackage{dcolumn}
\usepackage{bm}
\usepackage{subfigure}

\begin{document}

\title{Real-space observation of skyrmion clusters with mutually orthogonal skyrmion tubes}

\author{Hayley. R. O. Sohn}
\thanks{leonov@hiroshima-u.ac.jp}
\affiliation{Soft Materials Research Center and Materials Science and Engineering Program,
University of Colorado, Boulder, CO 80309, USA}

\author{Sergei M. Vlasov}
\affiliation{ITMO University, 197101, Saint Petersburg, Russia}

\author{Valeriy M. Uzdin}
\affiliation{ITMO University, 197101, Saint Petersburg, Russia}
\affiliation{Department of Physics, St. Petersburg State University, St. Petersburg 198504, Russia}

\author{Andrey O. Leonov}
\thanks{leonov@hiroshima-u.ac.jp}
\affiliation{Chirality Research Center, Hiroshima University, Higashi-Hiroshima, Hiroshima 739-8526, Japan}
\affiliation{Department of Chemistry, Faculty of Science, Hiroshima University Kagamiyama, Higashi Hiroshima, Hiroshima 739-8526, Japan}
\affiliation{IFW Dresden, Postfach 270016, D-01171 Dresden, Germany} 

\author{Ivan I. Smalyukh}
\thanks{ivan.smalyukh@colorado.edu}
\affiliation{Soft Materials Research Center and Materials Science and Engineering Program,
University of Colorado, Boulder, CO 80309, USA}
\affiliation{Department of Physics and Department of Electrical, Computer and Energy Engineering,
University of Colorado, Boulder, CO 80309, USA}
\affiliation{Renewable and Sustainable Energy Institute, National Renewable Energy Laboratory
and University of Colorado, Boulder, CO 80309, USA} 

\date{\today}

\begin{abstract}
{We report the discovery and direct visualization of skyrmion clusters with mutually-orthogonal orientations of constituent isolated skyrmions in  chiral liquid crystals and ferromagnets. 
%
%
We show that the nascent conical state underlies attracting inter-skyrmion potential, whereas an encompassing homogeneous state leads to the  repulsive skyrmion-skyrmion interaction. 
The crossover between different regimes of skyrmion interaction could be identified upon changing layer thickness and/or the surface anchoring. 
We develop a phenomenological theory describing  two types of skyrmions and the underlying mechanisms of their interaction.
We show that isolated horizontal skyrmions with the same polarity may approach a vertical isolated skyrmion from both sides and thus constitute two energetically-different configurations which are also observed experimentally. 
In an extreme regime of mutual attraction, the skyrmions wind around each other forming compact superstructures with undulations.
%
%
%
%
We also indicate that our numerical simulations on skyrmion clusters are valid in a parameter range corresponding to the A-phase region of cubic helimagnets. 
}
\end{abstract}

         
\maketitle


\textit{Introduction.} 
%
%
Magnetic chiral skyrmions are particle-like topological solitons with complex non-coplanar spin structures \cite{JMMM94,Romming13,LeonovNJP16,review} stabilized in noncentrosymmetric magnetic materials by specific Dzyaloshinskii-Moriya interaction (DMI) \cite{Dz64}.
DMI provides a unique stabilization mechanism, protecting  skyrmions from radial instability \cite{JMMM94,LeonovNJP16} and overcoming the constraints of the Hobart-Derrick theorem \cite{solitons}.
Recently, skyrmion lattice states (SkL) and isolated skyrmions (ISs) were discovered in bulk crystals of chiral magnets near the magnetic
ordering temperatures (A-phase) \cite{Muehlbauer09,Wilhelm11,Kezsmarki15} and in thin nanolayers 
over larger temperature regions\cite{Yu10,Yu11,Du15,Liang15}. 
%
%
The energy contributions phenomenologically analogous to DMI also arise in chiral liquid crystals (LC) and are responsible for a surprisingly large diversity of naturally-occurring and laser-generated topologically nontrivial solitons (including skyrmions)  \cite{Smalyukh10,Ackerman17,Ackerman16}. 
%

\begin{figure*}
\includegraphics[width=1.99\columnwidth]{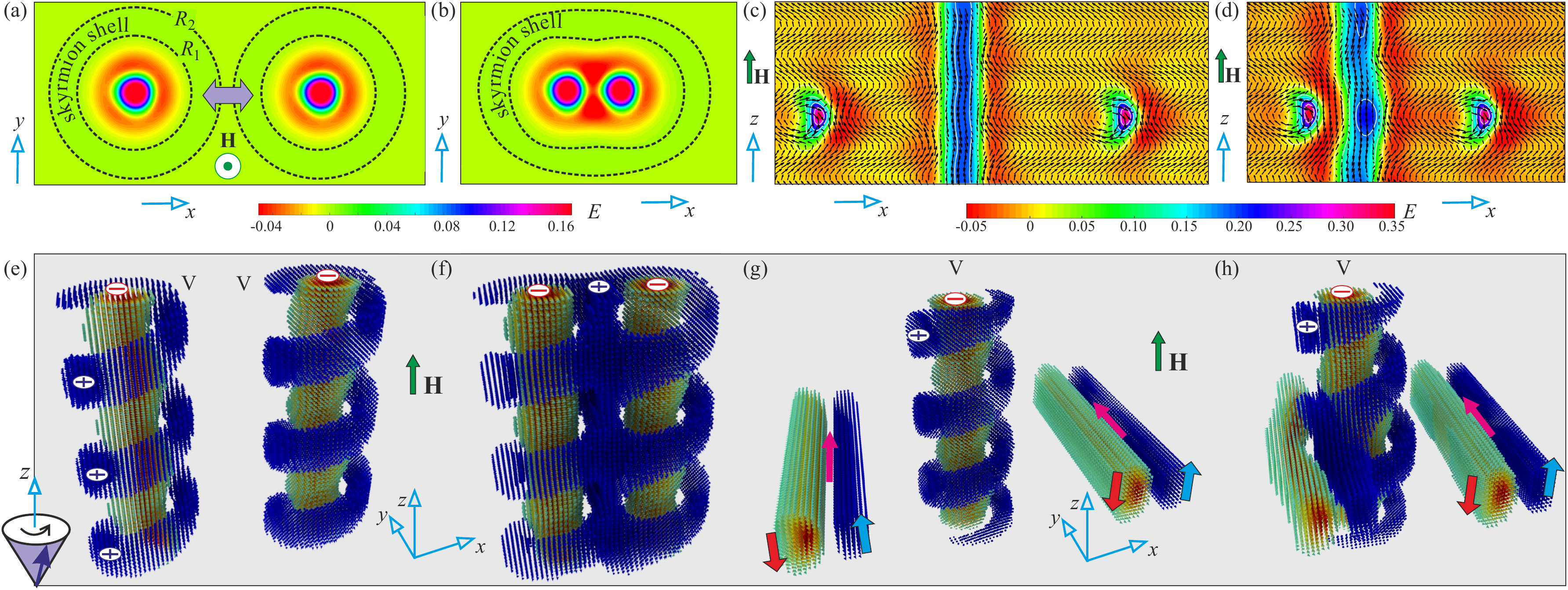}
\caption{(color online) Placed into the conical phase, each vertical IS marked by V  develops a transitional region (named shell) with positive energy density (radial region that stretches between radii $R_1$ and $R_2$ in (a)). Then, an attractive V-V interaction and skyrmion cluster formation (b) can be explained  by the overlap of shells and ensuing reduction of total energy. The color plots in (a) and (b) indicate the energy density averaged over $z$-coordinates and plotted on the $xy$ plane. 
The same "energetic" arguments can be applied to also address an attraction between vertical  and horizontal ISs (c), (d). 
The color plots in (c) and (d) indicate the energy density counted from the conical state and plotted on the $xz$ plane  for a cross section along the skyrmion centers. The black arrows are projections of the magnetization on to the $xz$ plane. $H/H_D=0.5$ for both cases.
The attracting nature of IS-IS interaction can alternatively be explained based on the principles of compact cluster formation (e) - (h) (see text for further details). Then, the cluster formed out of two vertical ISs (f) recreates the fragment of a metastable SkL with the homogeneous state between two skyrmions. Such a homogeneous state is formed if the coils of two skyrmions are getting "zipped" together. We extract the spins corresponding to the conical phase from the obtained spin configurations  what facilitates such way of reasoning. 
The magnetic field (green arrow) is applied along $z$. Pink arrows in (g) and (h) show the magnetization direction along $y$, i.e., the polarity of horizontal skyrmions. Red and blue arrows indicate the magnetization direction in the cores and transitional regions of horizontal skyrmions.
\label{art}}
\end{figure*}


The nanometer size of magnetic skyrmions, their topological protection and the ease with which they can be manipulated by electric currents \cite{Schulz12,Jonietz10,Hsu17} has opened a new active field of research in non-volatile memory and logic devices \cite{Sampaio13,Tomasello14}. 
In particular, in the skyrmion racetrack \cite{Tomasello14,Muller17,Wang16} -- a prominent model for future information technology --  information flow is encoded in the isolated skyrmions \cite{Fert2013} moving within a narrow strip. 
LC-skyrmions with a typical size of several micrometers can also be set in a low-voltage-driven motion with the precise control of both the direction and speed \cite{Ackerman17c}. 
The LC-skyrmions are confined in a glass cell with thickness comparable with the helicoidal pitch, an LC counterpart of a race-track memory. 

%
The complex three-dimensional internal structure of magnetic ISs and character of IS-IS 
interaction  are imposed  by a surrounding "parental" state, e.g.,  a state  homogeneously magnetized along the field (repulsive inter-skyrmion potential)\cite{LeonovNJP16}, a conical phase with the wave vector along the magnetic field (attraction) \cite{LeonovJPCM16,LeonovAPL16} or a tilted ferromagnetic state 
(anisotropic potential) \cite{Leonov17}. 
The same regimes of skyrmion interaction are experimentally observed in chiral liquid crystals. In particular, an analogue of the conical phase  can be achieved as a result of the competition between the director's tendency to twist and surface anchoring accompanied by the confinement at finite sample thickness, which could be further enriched, say, by applying the electric or magnetic fields. 
Whereas the surface anchoring tends to orient the director $\mathbf{n} (\mathbf{r})$ perpendicular to the confining glass plates, the electric field $\mathbf{E}$ in materials with  negative dielectric anisotropy $\Delta\varepsilon$ (analogous to the easy-plane anisotropy) alongside with  the director's tendency to twist due to the medium's chiral nature 
 -- tends to orient $\mathbf{n} (\mathbf{r})$ parallel to them \cite{Ackerman17c}. 
%


%

%

%


Additionally, magnetic (or LC) skyrmion tubes may orient 
either along or perpendicular to an applied magnetic field \cite{Leonov2018a,Sohn17} (to glass substrates). 
The first type of ISs (vertical skyrmions, marked by V in Fig. \ref{art}) perfectly blend into the homogeneously saturated state  whereas the second one (horizontal skyrmions)  -- into the spiral state. Then, a crossover between the two 
takes place for an intermediate value of an applied magnetic field (or carefully adjusted values of an electric field, surface anchoring, and film thickness) and enables complex cluster formation with mutually perpendicular arrangement of skyrmions \cite{Leonov2018a}. 
%
%
The aforementioned skyrmion traits  do not only open up new routes for manipulating these quasi-particles in energy-efficient spintronics applications, but also highlight a paramount role of cluster formation  within the A-phases of bulk cubic helimagnets near the ordering temperature (e.g., in B20 magnets MnSi \cite{Muehlbauer09} and FeGe \cite{Wilhelm11}), the fundamental problem that gave birth to skyrmionics. 
%


In the present paper, by experimental and numerical means, we formulate the principles 
of  cluster formation from isolated and mutually-perpendicular skyrmions. 
We show that horizontal ISs couple with the vertical ISs and form two experimentally-distinguishable configurations different in their pair energy and relative distance. 
The regimes of IS-IS interaction are switched according to the angle of the surrounding conical phase. 
At the verge  of the homogeneous state, the ISs develop repulsive interaction with distant mutual location.
On the contrary, with the increase of the conical angle, the ISs enter an extreme regime of attraction leading to their undulations.
Our choice of chiral liquid crystals to vindicate theoretical concepts of skyrmion cluster formation and to provide insights into similar structures in chiral magnets is stipulated by the obvious advantages of LCs over magnetic systems in terms of experimental testing: (i) the system parameters can be varied over wide limits to achieve a necessary regime of inter-skyrmion potential; (ii) as a rule, experiments are conducted at room temperature and are comparatively simple; (iii) the results of investigations are easily visualized by polarizing optical microscopy \cite{Ackerman17,Sohn17}, to a degree not usually attainable in the investigation of magnetic systems.
%

\textit{Phenomenological model. }
%
We consider the standard model for magnetic states in cubic non-centrosymmetric ferromagnets \cite{Dz64,Bak80},
\begin{eqnarray}
w =A\,(\mathbf{grad}\,\mathbf{m})^2 + 
D\,\mathbf{m}\cdot \mathrm{rot}\,\mathbf{m}
- \mu_0 M \mathbf{m}\cdot \mathbf{H},  
\label{density}
\end{eqnarray}
where $\mathbf{m}= (\sin\theta\cos\psi;\sin\theta\sin\psi;\cos\theta)$ is the unity vector along the magnetization $\mathbf{M}$, 
$A$ is the exchange stiffness constant, $D$ is the Dzyaloshinskii-Moriya coupling energy, and $\mathbf{H}$ is the applied magnetic field.
Chiral modulations along the applied field  with the period $L_D = 4 \pi A/|D|$ correspond to the global minimum of the functional (\ref{density}) below the critical  field $ \mu_0 H_D = D^2/(2 A M)$.
The equilibrium parameters for the cone phase are expressed in analytical form \cite{Bak80} as $\theta_c = \arccos \left( H/H_D \right) , \quad \psi_c = 2\pi z/L_D$  where $z$ is the spatial variable along the applied field.
%
%
At $H = H_D$,  the cone phase transforms into the saturated state with $\theta = 0$ and thus underlies the crossover between two regimes -- inter-skyrmion attraction and repulsion \cite{LeonovJPCM16,Loudon18}.
%

The first two terms in model (\ref{density}) also correspond to the elastic energy contributions in the Frank free energy 
that pertain to splay $K_1$, twist $K_2$ and bend $K_3$ distortions of the director provided that the one-constant approximation $K_1=K_2=K_3=K$ is utilized \cite{books}: $A \rightarrow K/2$, $D \rightarrow Kq_0$. Indeed, the values of elastic constants in common LCs are comparable and this one-constant approximation is commonly used.
The absence of a Zeeman-like term in LCs 
is compensated by the interplay of the electric field  and the surface anchoring. As a result, the LC conical phase acquires the angle varying across the layer thickness. 

\begin{figure}
\includegraphics[width=0.99\columnwidth]{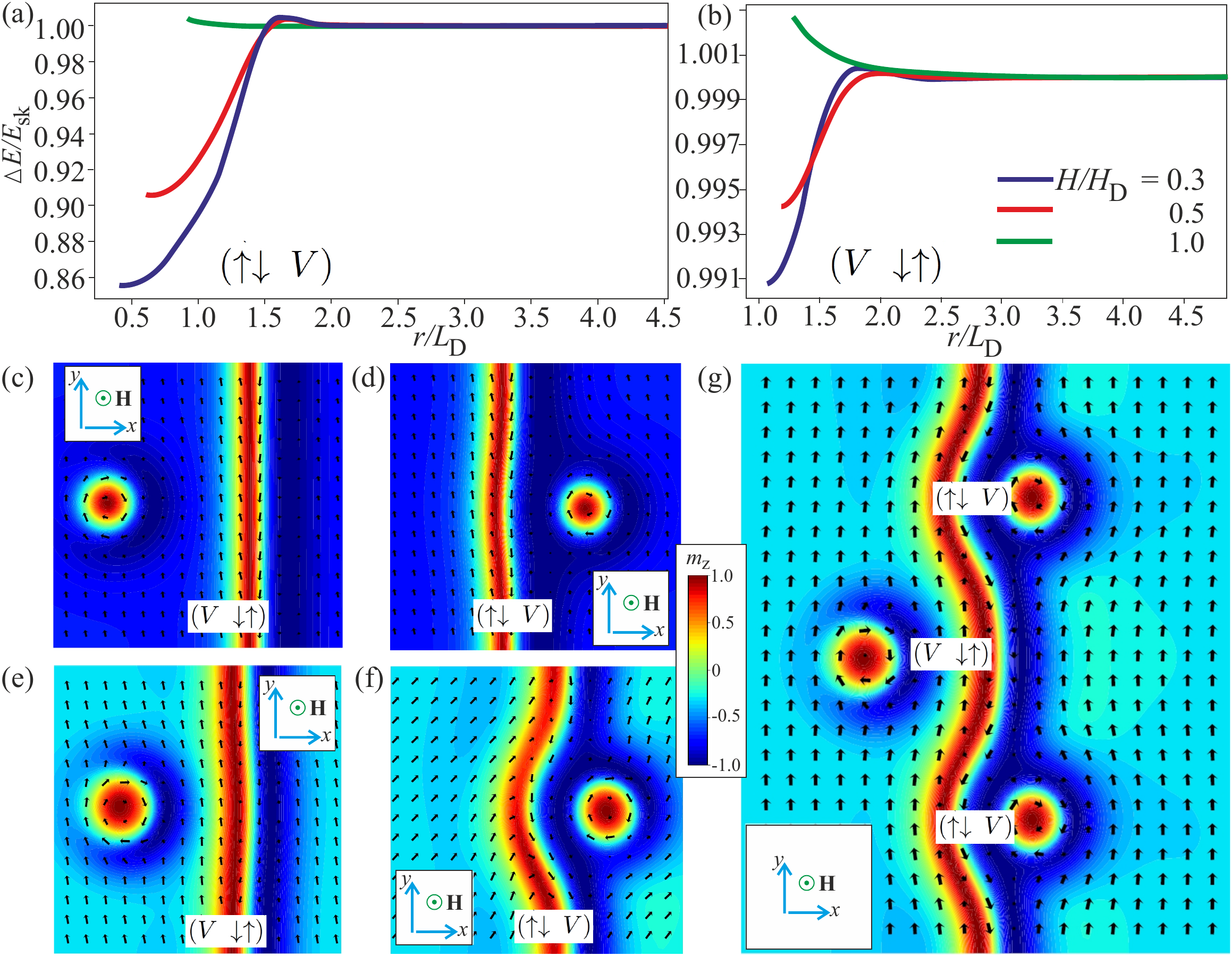}
\caption{ (color online) Reduced energy for the interaction between two ISs in the configuration $(\downarrow\uparrow\,V)$ (a) and $(V\,\downarrow\uparrow)$ (b) calculated as a function of the distance $r$ between the skyrmion centers for different values of the applied magnetic field. Numerical solutions for $(\downarrow\uparrow\,V)$  configurations ($h=0.7$ (d), $h=0.3$ (f)) show that for lower values of the fields a horizontal skyrmion is strongly bent by the vertical one (f). In the configuration $(V\,\downarrow\uparrow)$ (c), (d), no pronounced bending is observed. Thus, in the mixed configuration that includes both $(V\,\downarrow\uparrow)$ and $(\downarrow\uparrow\,V)$, the undulated structure of a horizontal skyrmion is only stipulated by the vertical skyrmions from the right side, whereas vertical skyrmions from the left already adjust to such a curved structure. 
\label{interaction}}
\end{figure}

\textit{The physical principles of skyrmion cluster formation.} Attraction of vertical skyrmions (Fig. \ref{art} (a), (b)) mediated by the conical phase was considered theoretically  in Refs. \onlinecite{LeonovJPCM16,LeonovAPL16,Ackerman17c}. Experimentally, clusters of such ISs have been observed in thin (70 nm) single-crystal samples of Cu$_2$OSeO$_3$ taken using transmission electron microscopy \cite{Loudon18} and in nematic fluids, where they were shown to also form skyrmion chains\cite{Ackerman17c}. 
The attracting nature of skyrmion-skyrmion interaction  
was explained on the basis of the so-called skyrmion shell -- 
a domain wall region separating the skyrmion core from the cone phase that after averaging over the $z$-coordinate has the positive energy density (region between two concentric dotted circles in Fig. \ref{art} (a)) \cite{LeonovJPCM16,LeonovAPL16,Inoue2018}. 
On the same 
principle, 
the mutual interaction among horizontal  and vertical ISs \cite{Leonov17}  
can be addressed (Fig. \ref{art} (c) - (d)): in general, both orientations of non-axisymmetric skyrmions are geometrically incompatible with their host cone phase what necessarily leads to the excessive amount of energy density. 
%

In the present manuscript, we adopt another approach that proves to also be instructive in the case  of horizontal skyrmions. 
After we have extracted the spin components corresponding to the conical phase, the vertical IS represents a cylinder-like core centered around the magnetization opposite to the field and a coil with the magnetization along the field (Fig. \ref{art} (e)). The  cluster formation may then be visualized as a process when the loops of a coil of one IS fill the voids between the loops of another IS (Fig. \ref{art} (f)). By this, the compact skyrmion pair recreates a fragment of a SkL that within the model (\ref{density}) is a metastable state: the magnetization on the way from the center of one IS to the center of another has a rotational fashion as in ordinary axisymmetric skyrmions and thus reduces the energy of a shell in the inter-skyrmion region.

Horizontal skyrmions within the same paradigm 
are displayed as two parallel distorted cylinders centered around the core (red) and the transitional region (blue) with the negative and the positive $m_z$-component of the magnetization, correspondingly (Fig. \ref{art} (g)). 
For consistency, we run horizontal skyrmions along the $y$-axis. Then, one could distinguish horizontal skyrmions with positive or negative polarity (direction of the magnetization in the region between two cylinders, with the pink arrow in Fig. \ref{art} (g)) being moved toward the vertical IS either with their core or a transitional region first. 
We accommodate the following notations for such skyrmions:  $(\downarrow\uparrow)$ (positive polarity) and  $(\uparrow\downarrow)$ (negative polarity) as read from left to right. 
Then, the equilibrium configurations depicted in Fig. \ref{art} (h) can be denoted as $(\downarrow\uparrow\,V)$ (horizontal IS with the positive polarity approaching the vertical IS from the left) and $(V\,\downarrow\uparrow)$ (from the right). 
In this notation, $(V\,\downarrow\uparrow)$ and $(\uparrow\downarrow\,V)$    are essentially the same configurations that differ, however, by  the $z$-coordinate of the horizontal ISs, which is imposed by the encompassing conical phase (Fig. \ref{art} (g)) and equals the spiral period. 
In the configurations of Fig. \ref{art} (g), (h) horizontal skyrmions are located at the same $z$-coordinate and form compact cluster states: 
%
%
 in the configuration $(\downarrow\uparrow\,V)$, the transient region of the $(\downarrow\uparrow)$ IS slides in-between the coils of the vertical skyrmion, whereas in the configuration $(V\,\downarrow\uparrow)$ the core of the $(\downarrow\uparrow)$ IS 
just touches the coil. 
%



\textit{Skyrmion interaction potentials.} Once  embedded in the cone phase, both configurations, $(\downarrow\uparrow\,V)$ and $(V\,\downarrow\uparrow)$,
 develop an attracting interaction. In Fig.  \ref{interaction} (a) and (b), we plot  the interaction energy 
 as a function of the distance $r$ between the centers  of horizontal and vertical skyrmions calculated for different values of the applied magnetic field using minimal energy path method \cite{Lobanov,Ivanov} (see the Supplemental Material on different  methods of interaction energy calculation). 
Such Lennard-Jones type potential profiles show that the attractive inter-skyrmion coupling is characterized by a low potential barrier
and a rather deep potential well establishing the equilibrium separation of skyrmions in the bound biskyrmion state. 
And experimentally-observed skyrmion configurations indicate both cases, i.e., skyrmions in the minimum of the interaction potential (the red oval in Fig. 3 (b)) and skyrmions unable to overcome that potential barrier (the blue oval in Fig. 3 (b)).

The largest interaction energy is achieved for  the configuration $(\downarrow\uparrow\,V)$  (Fig.  \ref{interaction} (a)) whereas for $(V\,\downarrow\uparrow)$ (Fig.  \ref{interaction} (b)) the energy profit due to the cluster formation is almost negligible. 
The skyrmions tend to produce more compact skyrmion "molecules" with increasing angle $\theta_c$ of the conical phase (decreasing magnetic field in Figs. \ref{interaction} (a), (b)). 
However, the equilibrium distance between ISs in $(V\,\downarrow\uparrow)$ (Fig. \ref{interaction} (c)) is essentially larger than that for the other pair (Fig. \ref{interaction} (d)).
In a regime of strong coupling, the vertical IS induces undulations of a horizontal IS  in  $(\downarrow\uparrow\,V)$ (Fig. \ref{interaction} (f)), but leaves the horizontal IS intact in $(V\,\downarrow\uparrow)$ (Fig. \ref{interaction} (e)).
%
%
In a case when several vertical ISs bend one  horizontal skyrmion in the configuration $(\downarrow\uparrow\,V)$, the vertical ISs on the other side  will simply occupy  the space between undulations, the structure also observed experimentally.

\begin{figure}
\includegraphics[width=0.99\columnwidth]{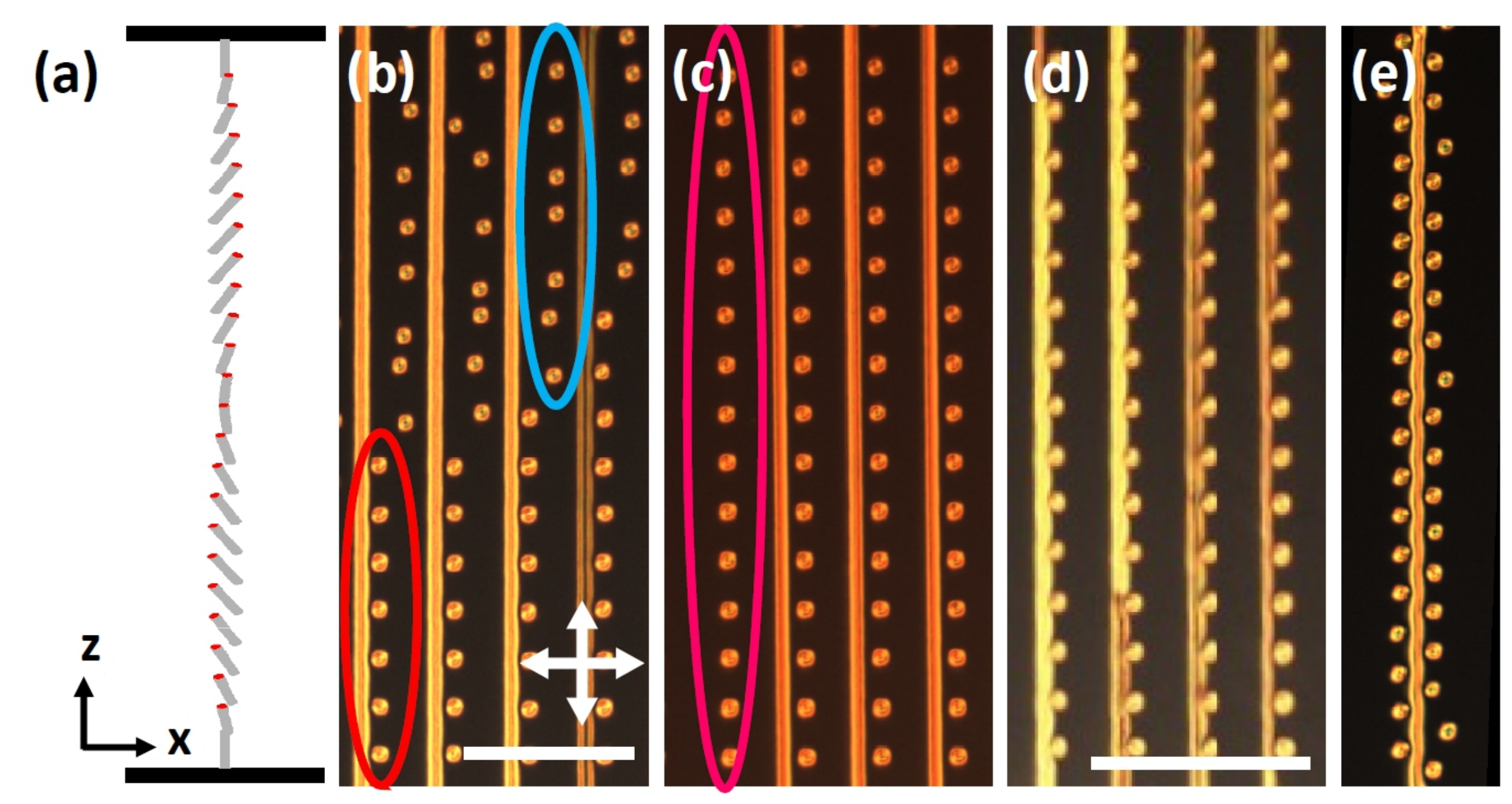}
\caption{ (color online) Experimental demonstration of orthogonal skyrmion interactions in a chiral nematic LC, visualized via polarizing optical microscopy images. Computer-simulated visualization of the tilted nematic liquid crystal director-field configuration (a) is depicted using cylinders with colored ends. The horizontal black bars represent the confining cell substrates. 
 Vertical skyrmions within the red oval in (b) correspond to the theoretical configuration $(\downarrow\uparrow\,V)$ (according to Fig. 2 (a)) and occupy the minimum of the interaction potential with the horizontal skyrmion located to the left. Vertical skyrmions in the blue oval in (b), however, have not overcome the potential barrier (Fig. 2 (a)) and are located at larger distances from the horizontal skyrmion to the left. Moreover according to Fig. 2 (b), these skyrmions are distant from the horizontal skyrmion to the right owing to their very weak coupling.
Such an anisotropic skyrmion interaction is additionally proven by the skyrmions in the pink oval of (c). These skyrmions are influenced only by the horizontal skyrmion from the right (configuration $(V\,\downarrow\uparrow)$) and are still located at larger distances. 
 Vertical skyrmions in a regime to induce undulations in the horizontal skyrmions (d) correspond to the theoretical configuration in  Fig. 2 (f). A horizontal skyrmion with an undulation induced by vertical skyrmions adherent to it (e)  corresponds to Fig. 2 (g). In polarizing optical microscopy, the crossed polarizer orientation is marked with white double arrows and the scale bars corresponding to 100 $\mu m$ are marked separately for (b-c) and (d-e) in (b) and (d), respectively. 
\label{experiment}}
\end{figure}

\textit{Experimental Methods.}
To prove the theoretical concepts, we prepared the chiral nematic LC mixtures by mixing a chiral additive with a room-temperature nematic host (see the Supplemental Material). 
To construct the LC cells with perpendicular surface boundary conditions and the cell gap $d = 10 \mu m$, glass substrates were treated with polyimide SE-1211 (purchased from Nissan). 
Both types of skyrmionic structure reported were controllably "drawn" in the LC cells using optical tweezers comprised of a 1064 nm Ytterbium-doped fiber laser (YLR-10-1064, IPG Photonics) and a phase-only spatial light modulator (P512-1064, Boulder Nonlinear Systems), as described in detail elsewhere \cite{Sohn17}.  
Because of the chiral LC’s preferences to twist, the twisted solitonic structures we selectively draw 
are energetically stable.

\textit{Experimental results.}
Although the cone angle effectively changes across the LC cell thickness (Fig. 3 (a)), this structure is analogous to that of the conical state of B20 magnets and can host both the horizontal and vertical skyrmions. 
To probe interactions between vertical and horizontal skyrmions, the vertical skyrmions were manipulated by the focused laser beams of tweezers. They were spatially translated within the sample plane and then released in the proximity of horizontal skyrmions, at different distances and on different sides of horizontal skyrmions. Depending on the release location, the skyrmions attracted (repelled) to (from) the horizontal skyrmions while undergoing Brownian motion, similar to that of colloidal particles \cite{Smalyukh07}.  This allowed us to establish the different inter-skyrmion interaction regimes discussed above.
At no applied fields, the background director is homeotropic at $d/p_0 < 1$ ($p_0 \approx  7.5-10 \mu m$ is helicoidal twist periodicity (pitch)) and has a translationally-invariant configuration at $d/p_0 = 1-1.35$, with the interactions between orthogonal skyrmions being repulsive in the former case, attractive in the latter case, and exhibiting richer behavior (dependent on position of the vertical skyrmion relative to the horizontal one) in the intermediate regime (Fig. 3).

\textit{Discussion and Conclusions.}
In the present manuscript, we supplement the energetic framework of the skyrmion interaction, according to which ISs overlap and form clusters to get rid of positive energy density, by the framework of cluster compactness. According to the formulated principles,  the surrounding  conical phase specifies not only the internal structure of both vertical and horizontal skyrmions, but also their relative orientation. Indeed, the horizontal skyrmions within the conical phase have only two possibilities to approach the vertical one -- either to fit between the coils or to touch it, both defined by the structure of the conical phase and by the tendency of compact cluster formation.   
%

On one hand, numerical calculations within the model (\ref{density}) are applicable for bulk helimagnets. 
Indeed, an internal structure of the surrounding conical phase was tuned by an applied magnetic field and the periodic boundary conditions in all three directions have been considered. 
In such a model, a crossover between two skyrmion orientations takes place for the same value of the field, at which the hexagonal SkL composed from the skyrmions along the field is believed to stabilize within the A-phase region of chiral B20 magnets (e.g., MnSi \cite{Muehlbauer09} and/or FeGe \cite{Wilhelm11}).
Moreover, a low-temperature SkL favored by the cubic anisotropy was recently observed experimentally  in the bulk insulator  Cu$_2$OSeO$_3$ \cite{Chacon}. 
%
%


On the other hand, the numerical results address the experimentally observed skyrmion behavior in  chiral liquid crystals. 
In spite of the fading intensity of (blue) skyrmion coils, caused by the surface anchoring, and just one spiral pitch that fits the layer thickness, the same principles of compact cluster formation as considered before may be validated. 
Then, skyrmion clusters from mutually orthogonal ISs may constitute a defect-free analogue of blue phases.
%
%
Thus, by employing analogies between the topology and energetics of liquid crystals and ferromagnets, we have provided experimental evidence for the rich behavior of skyrmionics structures within chiral condensed matter media, which we theoretically predict to exist in condensed matter systems such as B20 chiral magnets, chiral LCs and so on. Despite the different length-scales on which they occur, the interacting horizontal and vertical skyrmions in LCs exhibit behavior closely matching that of the analogous skyrmions in the non-centrosymmetric B20 magnets. Our study further reinforces the notion that  chiral LCs can be used as a model system for probing the behavior of skyrmionic structures on the mesoscopic scale. 

\textit{Acknowledgements. }
The authors are grateful to Jun-ichiro Ohe, Katsuya Inoue, Istvan Kezsmarki, and Maxim Mostovoy for useful discussions. This work was funded by JSPS Core-to-Core Program, Advanced Research Networks (Japan), JSPS Grant-in-Aid for Research Activity Start-up 17H06889, and Russian Foundation of Basic Research (grant RFBR 18-02-00267 A). AOL thanks Ulrike Nitzsche for
technical assistance.
The research at CU-Boulder  was supported by the US National Science Foundation through grants DMR-1810513 (research) and DGE-1144083 (Graduate Research Fellowship to H.R.O.S.).

\end{document}